\documentclass[12pt]{article}
\usepackage[T1]{fontenc}
\usepackage{stix2}
\usepackage{amsmath,graphicx}  
\usepackage[round]{natbib}     
\usepackage[hidelinks]{hyperref}
\newcommand{\code}[1]{\texttt{#1}} 

\usepackage{listings}
\lstdefinestyle{donuts}{language=R, basicstyle=\ttfamily\footnotesize,
columns=fullflexible, keepspaces=true, breaklines=true,
commentstyle=\itshape, frame=lines, numbers=none, showstringspaces=false,
upquote=true}
\usepackage{natbib}
\usepackage{setspace}
\usepackage[margin=1in]{geometry}

\begin{document}

\title{Oh doughnuts, it's the variance! A visual metaphor for the variance, standard deviation, and related quantities}
\author{Eugene D. Hahn \\
Department of Information Systems\\
and Operations Management\\
Salisbury University\\
Salisbury, MD 21801 USA}
\maketitle
  \begin{abstract}
    Despite the rise of artificial intelligence, human intuition endures as both a foundation for education and also (for now) a primarily human experience. We re-examine a formula for the sum of squares that is sometimes set aside because of possible numerical issues.  The formula can be used to create square doughnuts, a delicious regional favorite.  The variance becomes the dough, and we can examine the variance by either comparing doughnuts or by making different sizes of doughnuts with the same amount of dough.  In doing so, we provide a new visualization of a cornerstone formula, understandable by students at almost any level.  We further provide insight into related topics such as the coefficient of variation as well as a visualization of when numerical issues are likely to be problematic.  We also provide a new vocabulary for understanding the variance formula, with `dough' and `hole' taking the place of less edible concepts from the typical recipe. 
    
    \noindent \textbf{Keywords:} Area model; computational formula; standard deviation; statistics education; sum of squares; variance; visualization.
  \end{abstract}

\section{Introduction}
\label{sec:introduction}
The variance is an essential concept introduced in lower-level statistics classes.  Some essential concepts such as the mean are either familiar, or easy to support with visualizations, or both.  For example, it is easy to provide a visualization of the mean with reference to familiar objects such as seesaws, levers, or scales.
However students, and even some subject-matter experts \citep{OHagan98}, can find the variance to be less intuitively immediate.  In this article we bridge this gap by using doughnuts, an intuitively familiar sugary pastry, as a way of gaining a new perspective on the variance.

In the era of artificial intelligence, human intuition remains important for effective navigation of an increasingly complex world \citep{Vincent2021} and as a valuable asset for education \citep{Jagla1994}.  Visual representations are a well-established method for assisting with mathematical fluency \citep{Wilks1958}.  \citet[p.~xvi]{Polya1957} suggested drawing a figure as an important pre-requisite to solving mathematical challenges.  Memorable and humorous approaches can function as important teaching tools \citep{PearlLesser2022} and may increase student satisfaction \citep{LesserPearl2008}.
Square doughnuts are a regional favorite in the midwestern USA, and are just as tasty as round ones.  Square doughnuts also allow us to provide a new direct intuition about the nature of the variance.

\bigskip

Geometric interpretations of the variance and standard deviation have been provided by several authors.  \citet{Maverick1932} provided an approach for binned data.  He drew the sum of squared deviations as the area under a curve, allowing one to find the standard deviation.  \citet{PotterEtAl2010} presented the moments plot.  The moments plot takes inspiration from diagrams such as the box plot and the violin plot and facilitates the relative comparison of the first five moments in two or more datasets.  \citet{Orris2011} computed the squared deviations around the mean to produce squares.  These squares were rescaled into rectangles, which were combined to ultimately form a square representing the variance.  \citet{SarkarRashid2016} considered the empirical cumulative distribution function and presented an approach where the population standard deviation is the radius of a cylinder.  \citet{SarkarRashid2017} conceptualized the sample standard deviation as a three-dimensional volume of a right prism.  The mechanism for this approach arises from consideration of the hypotenuse of a triangle associated with a pair of deviation scores.  Most recently, \citet{Hanley2026} introduced modern visualizations to revisit Fisher’s groundbreaking proofs in terms of the $n$-dimensional geometry of sampling distribution concepts involving the standard deviation. Our doughnuts work at the other end of the educational curriculum, with a common formula students meet in their first statistics course.

\section{The Square Doughnut Construction: From Sums to Squares}
We begin with the sum of squared deviations around the mean, $\sum_{i=1}^n (x_i - \mu)^2$. Given data $x_1, \dots, x_n$, the population variance is
\begin{equation}\label{eq:varSSdev}
\sigma^2 = \frac{1}{n}\sum_{i=1}^n (x_i - \mu)^2.
\end{equation}
The population variance can also be written in terms of the raw sum of squares, $\sum_{i=1}^n x_i^2$:
\begin{equation}\label{eq:varSSdirect}
\sigma^2 = \frac{1}{n}\sum_{i=1}^n x_i^2 - \mu^2.
\end{equation}
Multiplying \eqref{eq:varSSdirect} by $n$ and re-arranging replaces averages with sums where all terms are re-scaled by the same factor.  We have:
\begin{equation}\label{eq:doughnutID}
 \sum_{i=1}^n x_i^2 \;=\; n\mu^2 + \sum_{i=1}^n (x_i - \mu)^2.
\end{equation}

We use \eqref{eq:doughnutID} as a key expression in our visualization. The core idea is simple. The sum of squared observations defines the exterior of a square region.  The term $n\mu^2$ corresponds to a concentric square hole. The dough that remains (the region minus the hole) is the sum of squared deviations $\sum_{i=1}^n (x_i - \mu)^2$. This nesting makes \eqref{eq:doughnutID} an actual difference in areas, and also reveals that the variance is the residual of raw moments.  We also can quickly obtain the sample variance, $s^2$, by the appropriate choice of denominator.  To minimize repetition we emphasize the population variance going forward with occasional detours into the sample variance and the population and sample standard deviations as needed.

Students are often advised to avoid \eqref{eq:varSSdirect}, and by extension \eqref{eq:doughnutID}, because of a computational caution.  In some cases \eqref{eq:varSSdirect} can suffer from catastrophic cancellation yielding decreased numerical accuracy \citep{Anscombe1967}.  This happens as the mean becomes increasingly large relative to the variance \citep{Welford1962,Ling1974,ChanGolubLeVeque1983}.  Students are therefore often (correctly) advised to use \eqref{eq:varSSdev} rather than \eqref{eq:varSSdirect}.  While \eqref{eq:doughnutID} is not always the best for numerical accuracy, it has strengths for teaching.  Our doughnuts make use of the following three quantities:
\begin{itemize}
    \item \textbf{Whole doughnut size:} area $= \sum x_i^2$,
    \item \textbf{Hole:} area $= n\mu^2$,
    \item \textbf{Dough:} area $= \sum(x_i - \mu)^2$.
\end{itemize}
We now have our ingredients for our new recipe.  The next thing is to bake.  As an analogy of baking the dough for a given number of minutes, we divide the dough.  Dividing the dough produces the final product, the variance.  So the population variance is easily remembered as $\textbf{Dough}/n$ while the sample variance is $\textbf{Dough}/(n-1)$.  We can think of the denominators as baking time and different choices of baking time produce slightly different doughnuts.

The sum-based identity
\begin{equation}\label{eq:sum-identity}
   \sum(x_i - \mu)^2 = \sum x_i^2 - n\mu^2
\end{equation}
reveals that the hole and dough partition the doughnut.  This identity allows for a quicker transition between $\sigma^2$ and $s^2$ by using the context of baking times as long as the instructor is comfortable using $\mu$ as a stand-in for $\bar{x}$.
We could also work with the mean-based identity
\begin{equation}\label{eq:mean-identity}
   \sigma^2 \equiv \frac{1}{n} \sum(x_i - \mu)^2  = \frac{1}{n}\sum x_i^2 - \mu^2.
\end{equation}
This identity is simpler for the purposes of explaining $\sigma^2$.  However this approach requires that $s^2$ be introduced in a slightly different way (and also requires the above change of $\mu$ to $\bar{x}$).   Yet another approach is that of \citet{Orris2011} who also uses the concept of squares to produce a (non-doughnut) visualization.  \citet{Orris2011} provided a development in terms of $s^2$ but not $\sigma^2$ and this choice of focus may correspond to instructor preference.  In that case our use of \eqref{eq:sum-identity} is also preferable over \eqref{eq:mean-identity}.

\section{Worked examples}

\subsection{Comparing the doughnut and the Berliner}
\begin{figure}[htb]
    \centering
    \includegraphics[scale=0.7]{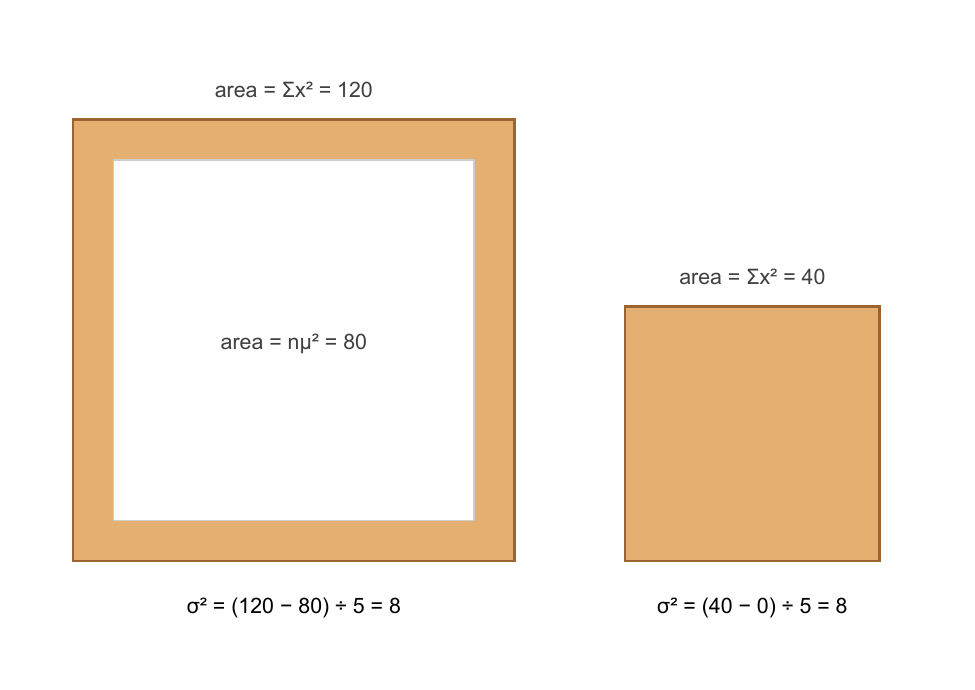}
    \caption{Square doughnut and Berliner for example data $x=\{1,2,3,5,9\}$ (left panel) and $x-4$ (right panel).}
    \label{fig:handout-doughnut}
\end{figure}

Consider a small dataset: $x = \{1, 2, 3, 5, 9\}$  with $n=5$.  We have $\mu = 4$, so $n\mu^2 = 5 \times 16 = 80$. Also $\sum x_i^2 = 1 + 4 + 9 + 25 + 81 = 120$. Then $\sum(x_i - \mu)^2 = 120 - 80 = 40$.
Figure~\ref{fig:handout-doughnut} shows the resulting visualization.  On the left, a brown band (dough) of area 40 surrounds a hole of area 80 inside a frame of area 120. Dividing the dough produces $\sigma^2 = 40/5 = 8$.   We have also placed a solid square doughnut produced from $x-4= \{-3, -2, -1, 1, 5\}$ beside it on the right.  A Berliner is a kind of doughnut that does not have a hole.  So the right-side dough square (with side length approximately 6.325) is a square Berliner that has the same amount of dough as the square doughnut in Figure \ref{fig:handout-doughnut}.  That is, $\sqrt{40} \approx 6.325$.  Again dividing the dough produces $\sigma^2 = 40/5 = 8$ for the Berliner.  We can now see that the hole vanishes when the mean of the data equals zero.  We can next refer back to \eqref{eq:varSSdirect} and see that when the data has mean zero, there is no need for concern about catastrophic cancellation because the subtractive step can be omitted.   This also provides context for multi-pass methods for computing the variance in which initially the mean is found and subtracted from the raw data \citep{Ling1974}.  When the mean is removed, either manually by mean-centering $x$ to $x-4$ as above or formulaically by a multi-pass calculation on $x$ as in \eqref{eq:varSSdev}, so too is the cause for numerical concern removed.

Taking square roots gives side lengths.  For example, $s_\text{outer} = \sqrt{\sum x_i^2}$ which is the length of the doughnut exterior.  Next $s_\text{inner} = \sqrt{n}\,|\mu|$ is the side length of the interior doughnut edge and also the side length of the hole.  Lastly $s_\text{dough} = \sqrt{\sum(x_i - \mu)^2}$ is the side length of the dough that has been reconstituted into a Berliner.  Thus $\sigma = s_\text{dough}/\sqrt{n}$.  With the current data we find $s_\text{outer} = \sqrt{120} \approx 10.954$, $s_\text{inner} = \sqrt{80} \approx 8.944$, and $s_\text{dough} = \sqrt{40} \approx 6.325$.

\subsection{Location shifts: Different doughnuts with the same dough}
Students often learn that the variance is unaffected by positive or negative location shifts in the data.  Here we provide intuition about this important property of the variance.  Below in Figure \ref{fig:shift} we subtract a constant of 2 from our example data set $x$.  We see that subtracting a constant from the dataset alters the mean without changing the variance.
\begin{figure}[htb]
  \centering
  \includegraphics[scale=0.7]{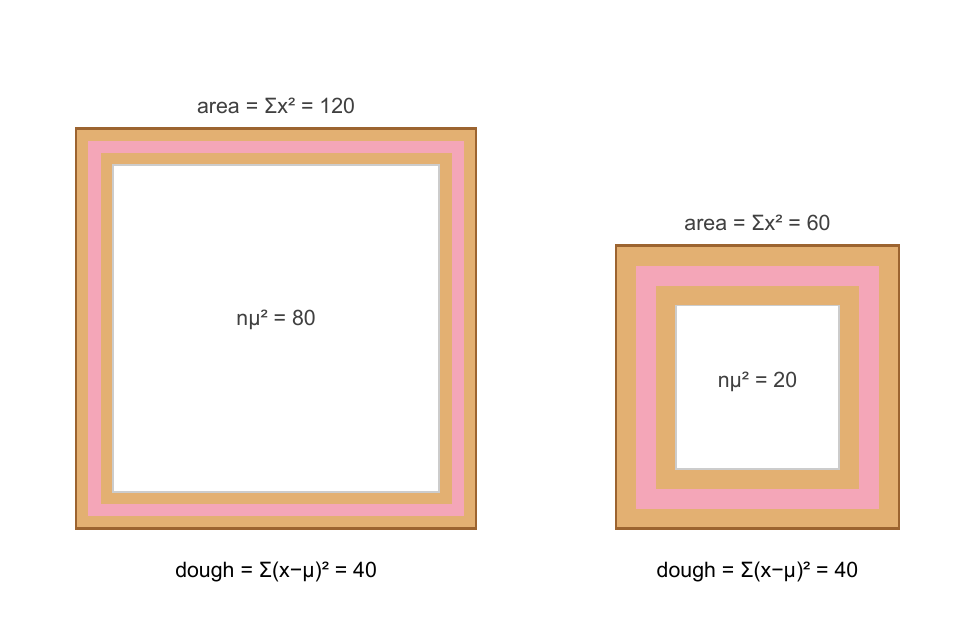}
  \caption{Same dough, different doughnuts: A comparison of $\{1,2,3,5,9\}$ and $\{1,2,3,5,9\}-2$.}
  \label{fig:shift}
\end{figure}
Although the two doughnuts in Figure \ref{fig:shift} have different sizes, they both have an equivalent amount of dough.  That means that both doughnuts in Figure \ref{fig:shift} would have the same amount of calories if they were actual doughnuts.  The concept of calories in the dough can be used to provide another metaphor for thinking about the variance.  We can also change our doughnuts in a decorative way to distinguish among concepts and examples or just to enhance appeal.  For example in Figure \ref{fig:shift} we have applied a decorative pink glazing to the doughnuts.

\subsection{Baking Times}

\begin{figure}[htb]
  \centering
  \includegraphics[scale=0.7]{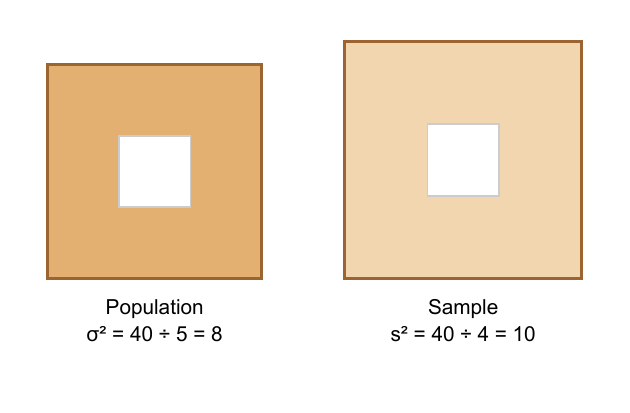}
  \caption{Baking the same dough, $\sum(x_i - \mu)^2=40$, for different times.}
  \label{fig:doughmode}
\end{figure}

One important educational outcome is ensuring that the concepts of population variance and sample variance are associated with the correct denominators of $n$ and $n-1$ respectively.  It is this instructor's experience that an explanation of the elimination of a degree of freedom can be slightly elusive to some students, particularly towards the end of a long class period.  By contrast, it is easy (and even amusing) for students to see and compare two doughnuts, one of which has been baked longer than the other.  A choice of a different color shade can further amplify the comparison of the doughnut that has been baked for an `additional minute' versus one which has not.  Figure \ref{fig:doughmode} displays such a comparison using a single data set.  In this data set the population and sample variance are displayed next to one another.  The population variance is given a deeper tone given its extra baking time.

In Figure \ref{fig:doughmode} we use dataset $x-3=\{-2, -1, 0, 2, 6\}$ which has mean of unity.  Both doughnuts have an equal quantity of dough, namely both have $\sum(x_i - \mu)^2=40$.  On the left we have the population variance while on the right we have the sample variance.  We can see the areas of both holes are the same: the hole is fixed by the data while only the dough bakes.  The same 40 units of dough becomes the population and sample variances when baked for $n$ versus $n-1$ minutes respectively.  Students with baking experience might recognize that the doughnut hole likely \emph{would} be affected by the choice of baking time in the real world.  Fortunately the purpose of the visualization is to provide intuition about properties of the variance instead of being an exact representation of thermodynamic properties of pastry.

\subsection{The coefficient of variation}

We can also easily introduce the coefficient of variation using our doughnuts.  We have $\mathrm{CV} = s_\text{dough} / s_\text{inner}$ where $\mathrm{CV}$ is the coefficient of variation.
Furthermore the squared coefficient of variation is simply $\textbf{Dough}/\textbf{Hole}$.  That is
\begin{equation*}
\mathrm{CV}^2 = \frac{\sum_{i=1}^n (x_i - \mu)^2}{n\mu^2} = \frac{\sigma^2}{\mu^2}.
\end{equation*}
The concept of $\textbf{Dough}/\textbf{Hole}$ will be familiar to students after they have seen the previous example of Figure \ref{fig:shift} where the amount of dough is held constant and the area of the hole is varied. Using our example dataset $x$ we find that the $\mathrm{CV} = \sqrt{1/2}$ and $\mathrm{CV}^2 = 1/2$.

\section{Computational Corner: When the Shortcut Formula Fails}

\begin{figure}[htb]
  \centering
  \includegraphics[width=0.8\textwidth]{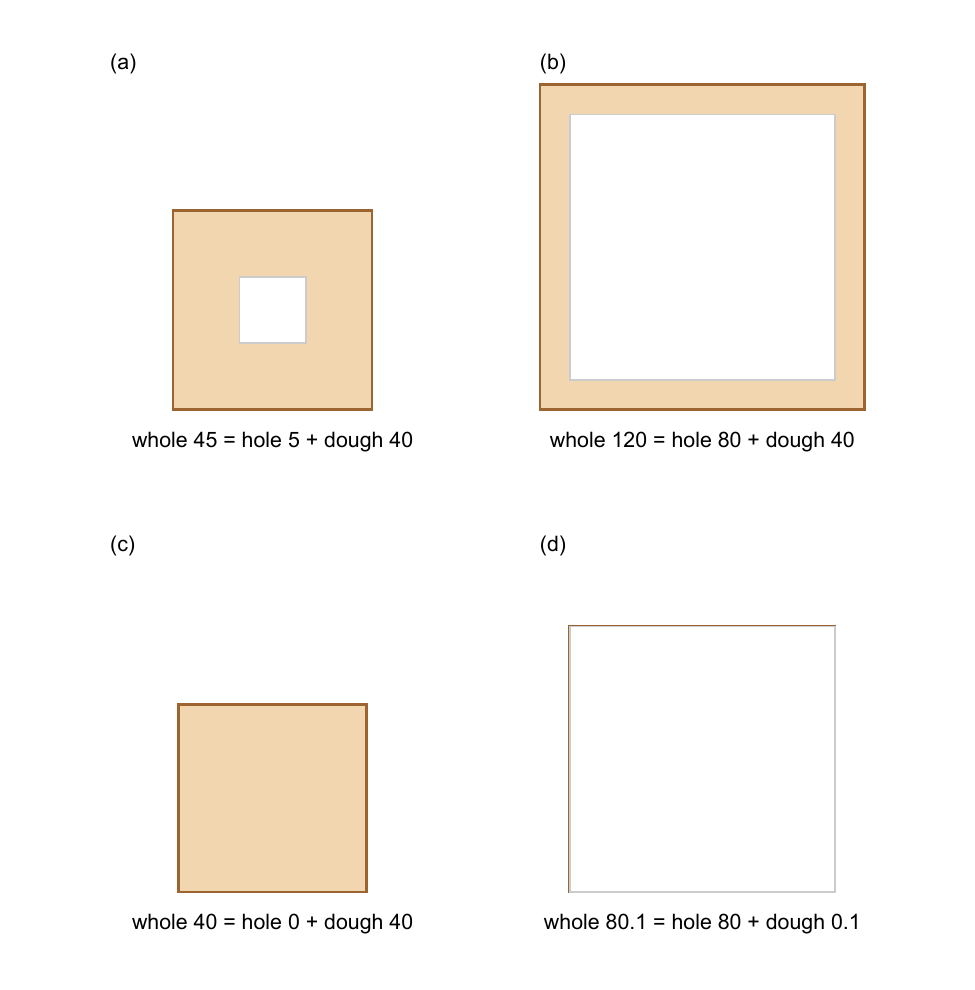}
  \caption{Four doughnuts drawn to the same scale: (a) $x-3=\{-2,-1,0,2,6\}$,
    (b) $x=\{1,2,3,5,9\}$, (c) $x-4=\{-3,-2,-1,1,5\}$, (d) $y=\{3.8,3.9,4.0,4.1,4.2\}$.
    Panels (a)--(c) have the same dough; (d) is glaze thin.}
  \label{fig:gallery}
\end{figure}

We mentioned earlier that a long-standing problem with the use of \eqref{eq:varSSdirect} is that it can lead to numerical problems resulting in inaccuracy.   We can now provide an elegant and straightforward explanation.  Simply stated, the formula \eqref{eq:varSSdirect} becomes increasingly suspect as the doughnut becomes more difficult to draw (and to see).  In more detail, as the mean grows relative to the variance, the hole gets larger and larger.  If the variance is fixed (by assumption), then a constant amount of dough must surround an increasingly large hole.  As the doughnut itself gets increasingly `glaze-thin', the formula of \eqref{eq:varSSdirect} becomes increasingly likely to be affected by catastrophic cancellation.   Figure \ref{fig:gallery} compares four doughnuts including one that is glaze thin in Panel (d).

The data corresponding to the respective doughnuts appears in Figure \ref{fig:gallery}'s caption.  Of these four doughnuts, the data underlying the glaze-thin one in Panel (d) is the most at-risk for numerical inaccuracy if \eqref{eq:varSSdirect} were used to compute the variance.  The reason why \eqref{eq:varSSdirect} fails is immediately obvious: the visual `resolution' has deteriorated and this doughnut is barely visible as an actual doughnut. So subtracting the entire hole from the glaze-thin dough is subject to loss of numerical accuracy. In a more advanced statistics class, we could rewrite the condition number $\kappa$ \citep[eq.~2.1]{ChanGolubLeVeque1983} in terms of concepts we have introduced above.  That is, $\kappa^2 = \sum x_i^2 / \sum(x_i-\mu)^2 = \textbf{Whole} / \textbf{Dough}$.  Then \eqref{eq:varSSdirect} becomes more subject to loss of numerical accuracy as $\mathrm{CV}$ goes to 0 and as $\kappa^2$ goes to infinity.  A discussion of the fixed computing precision of a computer could be introduced in this context \citep{altmanetal2004}.

\section{Classroom Tools}
\label{sec:tools}

We created the paper's figures with three brief R \citep{R2025} functions written using the ggplot2
package \citep{Wickham2016}. The functions \verb"donut_stats()", \verb"donut_frame()" and \verb"donut_compare()" are listed in Appendix A.  Importantly the variance is calculated using \eqref{eq:varSSdev}, so accuracy of numbers is preserved as best as possible even when the doughnut is glaze thin. Areas are programmed to be drawn to scale so that two doughnuts being compared share one scale.  Basic checks include a note appearing when one doughnut may be too small to see and a warning is given when two data sets have different numbers of observations.  Instructors and students alike can consider the use of color as a part of their use of the R software.  Real-life doughnuts have many colors due to choices of glazes and dough ingredients.  We include \verb"donut_schemes" with options for doughnut color selection and customization `to taste'.

We also provide a Microsoft Excel version for wider adoption by courses that do not use R. The Excel workbook runs in Excel 2010 or later and does not require macros. Interestingly Excel's own internal variance functions have a checkered past with documented numerical issues \citep[][pp.~12--15]{altmanetal2004}, so our workbook also computes the dough from squared deviations as in \eqref{eq:varSSdev}. A `Start' tab gives instructions, and then students can use one of several presets to begin with on the `Data' tab.  Students can also enter up to 200 values of their own in each of two data sets on the `Data' tab.  The workbook draws both square doughnuts to the same scale on a dedicated separate `Donuts' tab.  Toggling the `Show Donuts' switch to `No' on the `Donuts' tab hides the display of the doughnuts.  This can give students an opportunity to guess which data set has more variance and can be used as a simple interactive classroom activity.  Toggling the `Show Donuts' switch to `Yes' reveals the square doughnut images.  The data for the Presets is given on a final sheet and can be left as-is for most purposes.

\section{Discussion}

In this article we revisit a variance formula, \eqref{eq:varSSdirect}, that is usually avoided in lower-level statistics courses. Despite its possible numeric demerits, \eqref{eq:varSSdirect} can be used to create a simple and attractive teaching tool that is understandable by students at almost any level at or above high school/secondary school. We find this formula not only gives a new entertaining meaning to the variance, it also helps to visualize when \eqref{eq:varSSdev} is preferred, and provides benefits of intuition linkages about related concepts such as the population standard deviation ($\sigma = s_\text{dough}/\sqrt{n}$ in a Berliner) and the coefficient of variation.

We found square doughnuts to ultimately be a more intuitive concept for a modern educational curriculum than circular doughnuts.  Circular doughnuts are undoubtedly more familiar, but they would require additional data-processing steps to account for the side trip into the geometry of circles.  Another consideration with circular doughnuts is that the area of the circle requires the geometric constant $\pi$ while many introductory courses already use this Greek letter for the population proportion.  This dual usage of a single symbol might be cumbersome since geometric concepts, while well-suited to introductory statistics classes \citep[e.g.,][]{Wickens1995}, seem to be missing from many modern textbooks.

This article is designed to provide new opportunities for fostering statistical intuition and even engagement.  The identities of \eqref{eq:varSSdev} and \eqref{eq:varSSdirect} are classical but the computer visualization is new and should fit easily in contemporary statistics and data science curricula where either Excel or R is used as a computational environment. Natural extensions include overlaying the individual squared deviations on the dough, and drawing the doughnut about a reference point other than the mean, so that moving the hole connects directly to shifting the data.  Future research is also planned where the concept is extended to the analysis of variance.  In short, our paper offers a new vocabulary for the variance, with ``dough'' and ``hole'' taking the place of less edible concepts from the typical recipe.

\appendix
\section{R code for the square doughnut}
\label{app:code}
The R functions below draw the doughnut visualizations presented in this article. They require the ggplot2 package. The function \code{donut\_stats()} computes the dough
from squared deviations about the mean, so it stays accurate for glaze-thin data.
An automated test suite has performed checks on the functions, and a package version is in preparation.

\begin{lstlisting}[style=donuts]
## variance_donuts.R: square-donut pictures of the variance (code listing)
## Whole donut = sum(x^2), hole = n * mean^2, dough = sum((x - mean)^2).
## Requires ggplot2. The code is pure ASCII: symbols are written as \u escapes.
## Condensed from the project's tested reference code (donuts_v1.R).
library(ggplot2)

## Symbols used in labels
S2 <- "\u00b2"; SIG <- "\u03c3"; RT <- "\u221a"; SUM <- "\u03a3"
MU <- "\u03bc"; MINUS <- "\u2212"; DIV <- "\u00f7"

## Print a number without hiding small values (0.1 is not shown as 0)
fmt_num <- function(v) {
  if (abs(v - round(v)) < 1e-8 * max(1, abs(v)))
    return(format(round(v), scientific = FALSE))
  if (abs(v) >= 0.1) return(sprintf("%.2f", v))
  format(signif(v, 2), scientific = FALSE)
}

## Decimals needed so two nearly equal sides print differently
side_digits <- function(a, b, d = 2) {
  while (d < 6 && round(a, d) == round(b, d) && abs(a - b) > 1e-12) d <- d + 1
  d
}

## The numbers behind the picture. The dough is computed from squared
## deviations about the mean (two-pass), so it stays accurate even when the
## mean is huge. 
donut_stats <- function(x, scale = c("sum", "mean"), boundary_tol = 1e-3) {
  scale <- match.arg(scale)
  if (!is.numeric(x) || length(x) == 0)
    stop("`x` must be a non-empty numeric vector.")
  if (any(!is.finite(x)))
    stop("`x` contains NA, NaN or infinite values; remove them first.")
  n <- length(x); mu <- mean(x)
  var_pop <- mean((x - mu)^2)            # dough per observation
  ex2 <- mu^2 + var_pop                  # mean of x^2, without cancellation
  k <- if (scale == "mean") 1 else sqrt(n)
  kappa2 <- if (var_pop > 0) ex2 / var_pop else if (ex2 > 0) Inf else 1
  bound <- n * kappa2 * .Machine$double.eps
  list(n = n, mean = mu, E_x2 = ex2, sum_x2 = n * ex2,
       variance = var_pop, ss_dev = n * var_pop,
       side_outer = sqrt(ex2) * k, side_hole = abs(mu) * k,
       side_dough = sqrt(var_pop) * k,
       kappa = sqrt(kappa2), kappa2 = kappa2,
       shortcut_bound = bound, near_boundary = bound > boundary_tol)
}

## Areas and symbols for the chosen scale
lab_helper <- function(st, scale) {
  m <- scale == "mean"
  list(area_out = if (m) st$E_x2 else st$sum_x2,
       area_hol = if (m) st$mean^2 else st$n * st$mean^2,
       area_dgh = if (m) st$variance else st$ss_dev,
       sym_out = if (m) paste0("E[X", S2, "]") else paste0(SUM, "x", S2),
       sym_hol = if (m) paste0(MU, S2) else paste0("n", MU, S2),
       sym_dgh = if (m) paste0(SIG, S2)
                 else paste0(SUM, "(x", MINUS, MU, ")", S2))
}

## Colour schemes: choose with color = 1 to 6, or by name.
## The default is 3, golden dough with a pink stripe.
donut_schemes <- function() {
  data.frame(id = 1:6,
    name   = c("golden", "light", "golden_pink", "golden_white",
               "light_pink", "light_white"),
    fill   = rep(c("#E3B072", "#F2D6AF", "#E3B072", "#F2D6AF"),
                 c(1, 1, 2, 2)),
    edge   = rep(c("#9C6430", NA, "#9C6430", NA), c(1, 1, 2, 2)),
    stripe = c(NA, NA, "#F4A6B8", "#FFF2E4", "#F4A6B8", "#FFF2E4"),
    stringsAsFactors = FALSE)
}
resolve_scheme <- function(color = 3, fill_dough = NULL, edge_dough = NULL) {
  sch <- donut_schemes()
  i <- if (is.numeric(color)) match(color, sch$id) else match(color, sch$name)
  if (length(color) != 1 || is.na(i))
    stop("`color` must be one of 1-", nrow(sch), " or a name: ",
         paste(sch$name, collapse = ", "), ". See donut_schemes().",
         call. = FALSE)
  out <- as.list(sch[i, ])
  if (!is.null(fill_dough)) out$fill <- fill_dough
  if (!is.null(edge_dough)) out$edge <- edge_dough
  out
}

## A square of dough at (x0, y0) with side s; `inner` is the side of the
## hole drawn on top (0 if solid). A stripe is a ring through the middle
## third of the dough. The crust edge is drawn last so nothing covers it.
## The "white" stripe is cream (#FFF2E4) so it never merges with the hole.
## Re-rolled squares get stripe = FALSE: a ring there would look like a hole.
dough_square <- function(p, x0, y0, s, inner, sch, filled = TRUE,
                         stripe = TRUE) {
  sq <- function(p, a, fill, colour, lw)
    p + annotate("rect", xmin = x0 + a, xmax = x0 + s - a, ymin = y0 + a,
                 ymax = y0 + s - a, fill = fill, color = colour,
                 linewidth = lw)
  edge <- if (is.na(sch$edge)) NA else sch$edge
  if (filled) {
    p <- sq(p, 0, sch$fill, NA, 0)
    w <- (s - inner) / 2
    if (stripe && !is.na(sch$stripe) && w > 0)
      p <- sq(sq(p, w / 3, sch$stripe, NA, 0), 2 * w / 3, sch$fill, NA, 0)
  }
  sq(p, 0, NA, if (filled || !is.na(edge)) edge else "grey40", 0.6)
}

## One donut. dough_mode = FALSE draws outlines only. show_sigma adds the
## dough re-rolled into a square (its side is sigma on the mean scale).
## divide_dough (sum scale) divides the dough by n and by n - 1.
## show_sides labels the sides; show_identity prints the core identity.
donut_frame <- function(x, scale = c("sum", "mean"), dough_mode = TRUE,
                        color = 3, fill_dough = NULL, edge_dough = NULL,
                        show_sigma = TRUE, divide_dough = TRUE,
                        show_sides = TRUE, show_identity = FALSE) {
  scale <- match.arg(scale)
  sch <- resolve_scheme(color, fill_dough, edge_dough)
  st <- donut_stats(x, scale); lb <- lab_helper(st, scale)
  s <- st$side_outer; h <- st$side_hole; sig <- st$side_dough
  tsz <- 3.6; dg <- side_digits(s, h)
  p <- dough_square(ggplot(), 0, 0, s, h, sch, dough_mode)
  if (h > 1e-12) {                                   # the hole
    hc <- s / 2
    p <- p + annotate("rect", xmin = hc - h/2, xmax = hc + h/2,
                      ymin = hc - h/2, ymax = hc + h/2,
                      fill = "white", color = "grey80", linewidth = 0.4)
    if (h > 0.15 * s) {
      if (show_sides)
        p <- p + annotate("text", x = hc, y = hc + 0.03 * s, size = tsz,
                          color = "grey25", label = sprintf(
                          "side = %s%s = %.*f", RT, fmt_num(lb$area_hol),
                          dg, h))
      p <- p + annotate("text", x = hc, y = hc - (show_sides * 0.05) * s,
                        size = tsz, color = "grey25", label = sprintf(
                        "area = %s = %s", lb$sym_hol, fmt_num(lb$area_hol)))
    }
  }
  xmax <- s
  if (show_sigma && sig > 1e-12) {                   # dough mode square
    x0 <- s + 0.20 * s
    p <- dough_square(p, x0, 0, sig, 0, sch, dough_mode, stripe = FALSE) +
      annotate("text", x = x0 + sig/2, y = -0.08 * s, size = tsz,
               label = sprintf("%s = %s%s = %s",
                               if (scale == "mean") SIG else "side", RT,
                               fmt_num(lb$area_dgh), fmt_num(sig)))
    xmax <- max(x0 + sig, x0 + sig/2 + 0.22 * s)
  }
  if (show_sides)
    p <- p + annotate("text", x = s/2, y = 1.07 * s, size = tsz,
                      label = sprintf("side = %s%s = %.*f", RT,
                                      fmt_num(lb$area_out), dg, s))
  p <- p + annotate("text", x = s/2, y = (1.07 + show_sides * 0.08) * s,
                    size = tsz, color = "grey25", label = sprintf(
                    "area = %s = %s", lb$sym_out, fmt_num(lb$area_out)))
  if (show_identity)                                  # the core identity
    p <- p + annotate("text", x = s/2, y = -0.10 * s, size = tsz,
                      label = sprintf("%s = %s + %s :  %s = %s + %s",
                      lb$sym_out, lb$sym_hol, lb$sym_dgh, fmt_num(lb$area_out),
                      fmt_num(lb$area_hol), fmt_num(lb$area_dgh)))
  ymin <- -0.22 * s
  if (divide_dough && scale == "sum") {               # divide the dough
    d <- lb$area_dgh; n <- st$n
    divide <- sprintf("Divide the dough:  %s%s = %s %s %d = %s (population)",
                     SIG, S2, fmt_num(d), DIV, n, fmt_num(d / n))
    if (n > 1)
      divide <- sprintf("%s  \u00b7  s%s = %s %s %d = %s (sample)", divide,
                       S2, fmt_num(d), DIV, n - 1, fmt_num(d / (n - 1)))
    p <- p + annotate("text", x = xmax / 2, y = -0.19 * s, label = divide,
                      size = tsz, color = "grey25")
    ymin <- -0.26 * s
  }
  p + scale_x_continuous(limits = c(-0.05 * s, xmax + 0.05 * s)) +
    scale_y_continuous(limits = c(ymin, 1.20 * s)) +
    coord_fixed(clip = "off") + theme_void(base_size = 12)
}

## Two donuts on one scale
## Safeguards: an internal area check, a warning when n differs on the sum
## scale, and a note when one donut is under 1% of the other's area.
donut_compare <- function(x, y, labels = c("Dataset 1", "Dataset 2"),
                          scale = c("sum", "mean"), dough_mode = TRUE,
                          show_sigma = FALSE, color = 3,
                          fill_dough = NULL, edge_dough = NULL,
                          show_names = TRUE, show_verdict = TRUE) {
  scale <- match.arg(scale)
  sch <- resolve_scheme(color, fill_dough, edge_dough)
  st1 <- donut_stats(x, scale); lb1 <- lab_helper(st1, scale)
  st2 <- donut_stats(y, scale); lb2 <- lab_helper(st2, scale)
  a1 <- lb1$area_out; a2 <- lb2$area_out
  if (max(a1, a2) == 0)
    stop("Both datasets are all zeros, so there is no donut to draw.",
         call. = FALSE)
  ratio <- min(a1, a2) / max(a1, a2)
  for (pr in list(list(st1, lb1), list(st2, lb2)))   # drawn = stated areas
    if (!isTRUE(all.equal(pr[[1]]$side_outer^2, pr[[2]]$area_out,
                          tolerance = 1e-8)) ||
        !isTRUE(all.equal(pr[[1]]$side_hole^2 + pr[[1]]$side_dough^2,
                          pr[[1]]$side_outer^2, tolerance = 1e-8)))
      stop("internal error: drawn areas do not match the statistics")
  if (scale == "sum" && st1$n != st2$n)
    warning("Datasets have different n (", st1$n, " vs ", st2$n, "). On ",
            "the sum scale, donut size also grows with n; consider ",
            "scale = \"mean\". The verdict compares variances.", call. = FALSE)
  if (ratio < 0.01)
    message(sprintf(paste0("Note: the smaller donut is %.2g%% of the larger ",
                           "one's area. It is drawn to scale and may be hard ",
                           "to see."), 100 * ratio))
  big <- list(st = st1, lb = lb1, nm = labels[1])    # larger donut on the left
  sml <- list(st = st2, lb = lb2, nm = labels[2])
  if (a2 > a1) { tmp <- big; big <- sml; sml <- tmp }
  sb <- big$st$side_outer; ss <- sml$st$side_outer
  x2 <- sb + 0.35 * sb; tsz <- 3.6
  add_one <- function(p, d, xoff) {
    s <- d$st$side_outer; h <- d$st$side_hole; sig <- d$st$side_dough
    p <- dough_square(p, xoff, 0, s, h, sch, dough_mode)
    if (h > 1e-12) {
      hc <- xoff + s / 2
      p <- p + annotate("rect", xmin = hc - h/2, xmax = hc + h/2,
                        ymin = s/2 - h/2, ymax = s/2 + h/2, fill = "white",
                        color = "grey80", linewidth = 0.4)
      if (h > 0.15 * s && s > 0.25 * sb)          # no label on a tiny donut
        p <- p + annotate("text", x = hc, y = s/2 + 0.03 * s, size = tsz,
                          color = "grey25", label = sprintf("%s = %s",
                          d$lb$sym_hol, fmt_num(d$lb$area_hol)))
    }
    if (show_sigma && sig > 1e-12)
      p <- dough_square(p, xoff + 1.15 * s, 0, sig, 0, sch, dough_mode,
                        stripe = FALSE)
    if (show_names)
      p <- p + annotate("text", x = xoff + s/2, y = s + 0.06 * sb,
                        label = d$nm, size = tsz + 0.4)
    p +
      annotate("text", x = xoff + s/2, y = s + (0.06 + show_names * 0.11) *
               sb, size = tsz, color = "grey25", label = sprintf(
               "area = %s = %s", d$lb$sym_out, fmt_num(d$lb$area_out))) +
      annotate("text", x = xoff + s/2, y = -0.10 * sb, size = tsz,
               label = sprintf("dough = %s = %s", d$lb$sym_dgh,
                               fmt_num(d$lb$area_dgh)))
  }
  p <- add_one(add_one(ggplot(), big, 0), sml, x2)
  va <- big$st$variance; vb <- sml$st$variance
  same <- isTRUE(all.equal(va, vb, tolerance = 1e-8)) || max(va, vb) < 1e-12
  same_n <- big$st$n == sml$st$n
  same_hole <- isTRUE(all.equal(big$lb$area_hol, sml$lb$area_hol,
                                tolerance = 1e-8))
  verdict <- if (same && same_hole && same_n) {
    "Same donut"                                  # identical pictures
  } else if (same && (same_n || scale == "mean")) {
    "Same dough, different donuts"
  } else if (same) {
    "Same variance, different donuts"
  } else if (same_n) {
    da <- big$lb$area_dgh; db <- sml$lb$area_dgh
    sprintf("More dough = more variance:  %s vs %s",
            fmt_num(max(da, db)), fmt_num(min(da, db)))
  } else {
    sprintf("Variance per observation:  %s %s vs %s %s",
            big$nm, fmt_num(va), sml$nm, fmt_num(vb))
  }
  xr <- x2 + max(ss, 0.5 * sb) + 0.05 * sb       # room for a tiny donut
  if (show_verdict)
    p <- p + annotate("text", x = (xr - 0.05 * sb) / 2, y = -0.25 * sb,
                      label = verdict, size = tsz + 0.6, fontface = "italic")
  out <- p +
    scale_x_continuous(limits = c(-0.05 * sb, xr)) +
    scale_y_continuous(limits = c((if (show_verdict) -0.35 else -0.18) * sb,
                                  1.25 * sb)) +
    coord_fixed(clip = "off") + theme_void(base_size = 12)
  attr(out, "verdict") <- verdict                 # kept even when not drawn
  attr(out, "stats") <- list(first = st1, second = st2, ss_ratio = ratio,
                             side_ratio = sqrt(ratio))
  out
}
\end{lstlisting}

\bibliographystyle{abbrvnat}

\bibliography{variance-donutsv3}

\end{document}